\documentclass[sigconf,screen]{acmart}
\usepackage{tabularray}
\usepackage{svg}
\usepackage{fontawesome}
\AtBeginDocument{%
  }

\acmDOI{XXXXXXX.XXXXXXX}

\acmISBN{978-1-4503-XXXX-X/2018/06}

\acmYear{2026}\copyrightyear{2026}
\acmConference[FSE Companion '26]{34th ACM Joint European Software Engineering Conference and Symposium on the Foundations of Software Engineering}{July 5--9, 2026}{Montreal, QC, Canada}
\acmBooktitle{34th ACM Joint European Software Engineering Conference and Symposium on the Foundations of Software Engineering (FSE Companion '26), July 5--9, 2026, Montreal, QC, Canada}

\definecolor{Silver}{rgb}{0.95,0.95,0.95}
\definecolor{Mercury}{rgb}{0.95,0.95,0.95}
\definecolor{Gallery}{rgb}{0.921,0.921,0.921}

\begin{document}


\title[A Preliminary Reliance-Control Framework for AI in Software Engineering]{Towards an Appropriate Level of Reliance on AI: A Preliminary Reliance-Control Framework for AI in Software Engineering}

\author{Samuel Ferino}
\email{samuellucas.demouraferino@monash.edu}
\orcid{0000-0002-5484-1169}
\authornotemark[1]
\affiliation{%
  \institution{Monash University}
  \city{Melbourne}
  \country{Australia}
}

\author{Rashina Hoda}
\orcid{0000-0001-5147-8096}
\email{rashina.hoda@monash.edu}
\affiliation{%
  \institution{Monash University}
  \city{Melbourne}
  \country{Australia}
}

\author{John Grundy}
\orcid{0000-0003-4928-7076}
\email{john.grundy@monash.edu}
\affiliation{%
  \institution{Monash University}
  \city{Melbourne}
  \country{Australia}
}

\author{Christoph Treude}
\email{ctreude@smu.edu.sg}
\orcid{0000-0002-6919-2149}
\affiliation{%
  \institution{Singapore Management University}
  \city{Singapore City}
  \country{Singapore}}
\renewcommand{\shortauthors}{Ferino et al.}

\begin{abstract}
How software developers interact with Artificial Intelligence (AI)-powered tools, including Large Language Models (LLMs), plays a vital role in how these AI-powered tools impact them.  While overreliance on AI may lead to long-term negative consequences (e.g., atrophy of critical thinking skills);  underreliance might deprive software developers of potential gains in productivity and quality. Based on twenty-two interviews with software developers on using LLMs for software development, we propose a preliminary reliance-control framework where the level of control can be used as a way to identify AI overreliance and underreliance. We also use it to recommend future research to further explore the different control levels supported by the current and emergent LLM-driven tools. Our paper contributes to the emerging discourse on AI overreliance and provides an understanding of the appropriate degree of reliance as essential to developers making the most of these powerful technologies. Our findings can help practitioners, educators, and policymakers promote responsible and effective use of AI tools.
\end{abstract}



\maketitle

\section{Introduction}

AI-powered tool usage, especially  LLM-based tools, has greatly risen in SE in recent years with capabilities supporting many different software development tasks e.g., code generation, test generation, and debugging \cite{hou:2024}. Benefits of adopting AI-powered tools include reduced software developers' efforts and saved development time \cite{meem:2025}. Weber et al. \cite{weber:2024} identified significant gains in productivity through their within-subject study with 24 software developers using the SPACE framework.  Attracted by the potential advantages, enterprise reports show an increasing interest from companies, such as McKinsey \cite{mayer:2025} and DORA \cite{dora:2025}. However, incidents like the one involving Replit-agent \cite{techco:2025}, which mistakenly deleted a live database and made \textit{``unauthorised changes to live infrastructure, wiping out data for more than 1,200 executives and over 1,190 companies"}, show the risks of increasing reliance on AI tools.

Excessive or unnecessary trust or dependence on AI can result in potential negative effects, as explored in recent studies \cite{ferino:2025}. For instance, Kosmyna et al. \cite{kosmyna:2025} conducted an experiment exploring the neural and behavioural consequences of LLMs involving 54 software engineers, research scientists, and students. Combining electroencephalography -- which records the electrical activity of the brain -- to assess cognitive load with Natural Language Processing and human and AI judges over four months, they identified AI users persistently underperforming at neural, linguistic, and behavioural levels. Overreliance on AI can also affect team dynamics, reducing team collaboration and communication \cite{khojah:2024}. Khojah et al. \cite{khojah:2024} found in their observational study with twenty-four software engineers a reduction in team communication for critical questions,  where team members may spend considerable time adjusting prompts instead of discussing the issues directly.

Because of this, there is a growing interest from the research and practice communities to investigate \textit{appropriate reliance} on AI (e.g.,  \cite{bo:2025, he:2023}), the middle ground between overreliance and underreliance. For instance, Collins et al. \cite{collins:2024} experimented with introducing different types of frictions, which are design elements with the purpose of increasing users' time, effort, or cognitive load of accessing an AI-generated output, as an approach to modulate overreliance. During their user study with 100 participants, they could reduce users' reliance, but those frictions resulted in a drop in click rate as an unintended effect. Despite researchers' efforts, Bo et al. \cite{bo:2025} argue that \textsl{"the question of how to induce appropriate reliance is still open"}. Appropriate reliance is described as \textit{"relying on the AI when it's correct, and relying on yourself when it's not"} \cite{bo:2025}.

This paper addresses the problem of achieving an appropriate level of reliance on AI by proposing a preliminary reliance-control framework for AI. To the best of our knowledge, this is the first reliance-control framework proposition for AI-supported SE work. Leveraging this relationship, our study focuses on guiding software developers and software team leaders towards an appropriate level of reliance on AI. Based on our analysis, we extend the definition beyond accuracy and define \textbf{\textit{Appropriate Level of Reliance on AI}} more broadly as,  \textbf{\textit{the extent to which one should rely on AI tools}}, encompassing different \textbf{\textit{levels of reliance}}, from \textit{self-reliance} to full reliance on AI (i.e., \textit{full automation}) and different \textbf{\textit{levels of control}} from \textit{self-control} to \textit{losing control over AI}. The contributions of this paper are:
\begin{itemize}
    \item Characterisation of reliance on AI and control over AI;
    \item A preliminary reliance-control framework of AI for SE;
    \item A boarder, more nuanced definition of appropriate reliance on AI in SE.
\end{itemize}

\begin{table*}
\centering
\footnotesize
\caption{Characterisation of Control over AI.}
\label{tab:levelsOfControl}
\begin{tblr}{
  width = \linewidth,
  colspec = {Q[150]Q[458]Q[145]},
  row{odd} = {Silver},
  hlines,
}
\textbf{Type of Control}       & \textbf{Description}                                                                                                                                                                                                     & \textbf{Example}                                   \\
Self-Control over AI           & Software developers do not just hold control over AI, but also over themselves. They can overcome any urge to use AI.                                                                                               & Abstaining from using AI  tools when studying                 \\
Taking Control over AI         & Software developers assume strong control over AI, aware of potential flaws in AI (e.g.,  data breaches). ~Because of this, developers are unable to fully experience the benefits offered by AI tools. & Prompting with dummy data due to privacy concerns  \\
Balanced Control over AI       & Software developers find the middle ground in terms of control over LLMs~leveraging LLM's automation capabilities to improve their software development speed, while mitigating losses.                               & Boilerplate code generation                        \\
Handing over the Control to AI & Software developers start to step back, allowing AI to lead development tasks (e.g., problem solving).                                                                                                                & Vibe coding                                        \\
Losing Control over AI         & Software developers lack in understanding of the code's structure and flow, which impacts their debugging capabilities.                                                                                                & AI tools adding bugs or vulnerabilities into generated code 
\end{tblr}
\end{table*}

\begin{table*}
\footnotesize
\centering
\caption{Characterisation of Reliance on AI.}
\label{tab:degreesReliance}
\begin{tblr}{
  width = \linewidth,
  colspec = {Q[110]Q[408]Q[200]},
  row{odd} = {Mercury},
  hlines,
}
\textbf{Type of Reliance} & \textbf{Description}                                                                                                                   & \textbf{Example}                                        \\
Self-Reliance             & Software developers having confidence in their development skills, not constantly needing to seek support to finish tasks.             & Turning off GitHub Copilot and coding by yourself        \\
Reliance on Colleagues    & Software developers start to increase reliance on LLMs; however, they still prioritise human support or collaboration to finish tasks. & Seeking colleagues for software architecture design \\
Appropriate Reliance      & Those software developers increase their use of LLMs. However, they are also constantly vigilant of potential LLM pitfalls.            & Cross-checking references                               \\
Overreliance on AI      & software developers having excessive trust or dependence towards LLM. As a result, they can feel an urge to use LLMs.                  & Prompting for brainstorming before trying themselves    \\
Full Automation                & Software developers who fully rely on LLMs' capabilities, enabling them to move their focus to other tasks                             & Meeting summarisation                                   
\end{tblr}
\end{table*}

\section{Related Work} \label{sec:relatedWork}

Ma et al. \cite{ma:2023} investigate methods to leverage human-AI capabilities to calibrate appropriate trust in AI-assisted decision-making. They developed a human decision-making model designed to predict possible decisions made by humans in similar tasks.  They evaluated the effectiveness of their model with two preliminary experiments, comparing it with human responses. Based on their findings, they proposed three exploitation strategies to leverage this model: \textit{direct display}, \textit{adaptive workflow}, and \textit{adaptive recommendation}. They conducted a between-subjects crowdsourcing experiment with 293 participants to evaluate those three strategies and found that the strategies promote more appropriate human trust in AI, while acknowledging that their model might not account for all edge cases.


Kim et al. \cite{kim:2025}  conducted a think-aloud study with 16 participants and a large-scale controlled experiment with 308 participants. They focused on identifying features of LLM responses that can promote appropriate reliance. They identified that explanations and sources play an important role in users' reliance, where the presence of explanations was correlated with an increase in reliance on correct and incorrect responses. At the same time, the presence of a source reduced the reliance on explanations that include inconsistencies.

Bo et al. \cite{bo:2025} conducted a randomised online experiment involving 400  participants who attempted two challenging tasks: LSAT logical reasoning and image-based numerical estimation. They aimed to evaluate three reliance interventions -- reliance disclaimer, uncertainty highlighting, and implicit answer -- as a way of achieving appropriate reliance. However, they found that those interventions fail to improve appropriate reliance.

While prior work has examined interventions to calibrate appropriate reliance on AI through explanations, uncertainty cues, or user interface design, these studies focus mainly on trust calibration at the interaction level. Our work differs by introducing a two-dimensional reliance-control framework that conceptualises how users’ degree of reliance co-evolves with their level of control over AI systems.

\begin{figure*}
  \centering
    \includegraphics[width=0.85\linewidth]{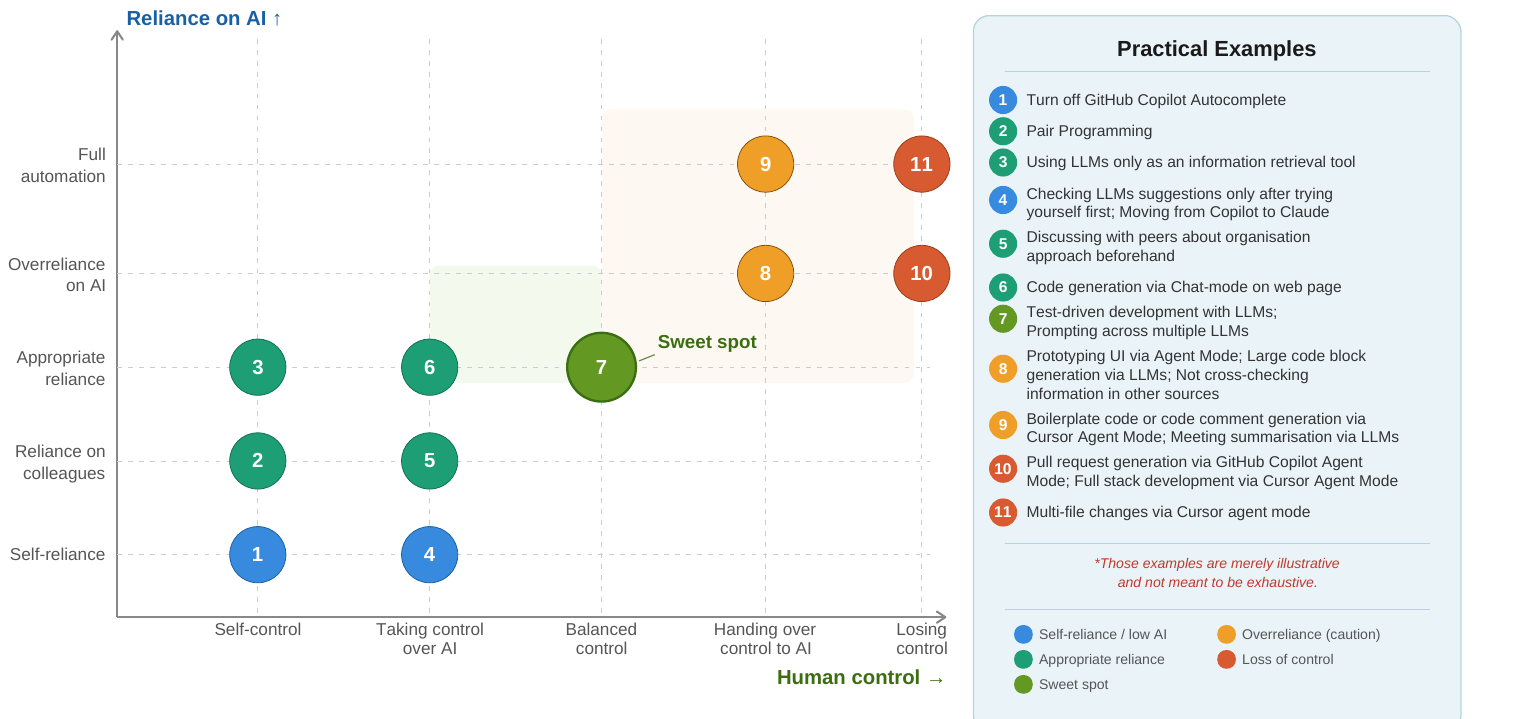}
  \caption{Reliance-Control Framework on AI for SE.}
  \Description{A coordinate graph, where y-axis represents the degree of reliance and x-axis represents the level of control.}
  \label{fig:framework}
\end{figure*}

\section{Reliance-Control Framework} \label{sec:preliminaryfindings}


\textbf{\textsc{Study Overview.}} We conducted twenty-two semi-structured interviews with software practitioners, in three rounds, focused on collecting their experiences of using LLM tools for software development. We employed socio-technical grounded theory (STGT) \cite{hoda2024book} for \textit{data analysis}. Our research protocol was approved by our Faculty Ethical Review Committee. The data analysis and collection took place over three rounds between October (2024) and September (2025), involving interview participants from Asian, Europe, North America, South America, and Oceania. Participants filled out a pre-interview questionnaire with their demographics and attitudes towards AI \cite{sindermann:2021}. During the interview, we discussed participants' experiences using LLMs. Participants used LLMs (e.g., Claude, ChatGPT) for several development tasks, such as coding and debugging. The first author performed open coding, constant comparison, and memoing. Before that, the first author and the second author, a grounded theory specialist, conducted open coding sessions with the first two interviews. During the analysis process, control and reliance emerged as prominent concepts. Based on the data analysis and collaborative brainstorming among the authors, we developed the framework diagram, which is illustrated in Figure \ref{fig:framework}. Further details are available in the online supplementary package \cite{researchArtifact}, including study protocol, interview guide, pre-interview questionnaire, researchers' assumption list, an example of how the data was analysed to derive the framework, and memos.

\textbf{\textsc{Framework Description.}} Table \ref{tab:levelsOfControl} presents the proposed degrees of control: \textit{self-control}, \textit{taking control}, \textit{balanced control}, \textit{handing over control}, and \textit{losing control}. Table \ref{tab:degreesReliance} characterises the proposed degrees of reliance: \textit{self-reliance}, \textit{reliance on colleagues}, \textit{appropriate reliance}, \textit{overreliance}, and \textit{full automation}. Both degrees of control and reliance may differ according to the software development task, as shown in the examples (last columns).

Figure \ref{fig:framework} shows our proposal of the relationship between reliance and control, with practical examples. Those examples are merely illustrative and not meant to be exhaustive. We envision the interception between \textit{balanced control} and \textit{appropriate reliance} as the optimal point (shown as point \textbf{\textcircled{7}}), often referred to as the sweet spot. In the remainder of this section, we now expand on the practical examples shown in the figure, all of which are grounded in real scenarios described by our interview participants.



\textbf{Self-Control and Self-Reliance \textcircled{1}} is exemplified when software developers turn off AI suggestions while familiarising themselves with new programming languages. A similar observation is made when SE students abstain from using LLMs by turning off GitHub Copilot suggestions until they solidify Computer Science fundamentals. \textbf{Self-Control and  Reliance on Colleagues \textcircled{2}} is illustrated when software developers prioritise traditional pair programming instead of AI pair programming. Pair programming sessions may be time-consuming, requiring two developers to work on a single task, and for this reason, they should be used to deal with complex problems. \textbf{Self-Control and Appropriate Reliance on AI \textcircled{3}} is evidenced when the software developer decides to use LLMs exclusively for information retrieval, for instance, replacing Google as the search engine. LLMs provide summarisation capabilities that reduce the effort of examining the different webpages resulting from traditional searches.

\textbf{Taking Control over AI and Self-Reliance \textcircled{4}} is exemplified when software developers only seek AI tools after trying them themselves. In this process, software developers may migrate from interactions inside the IDE to via the browser. For instance, migrating from Claude inside the Cursor to Claude in the online chatbot platform. This type of interaction with AI-powered tools forces the user to do context switching between the IDE and the LLM in the browser. \textbf{Taking Control over LLMs and Reliance on Colleagues \textcircled{5}} This occurs when software developers rely on discussion with colleagues to understand organisational approaches. In this situation, LLMs may provide suggestions, but may not align with the organisation's approach. 


\textbf{Balanced Control and Appropriate Reliance \textcircled{7}} is represented when software developers engage in test-driven development  (TDD) by leveraging agent mode capabilities of large language models found in AI-assistant development tools like Cursor, Windsurf, and GitHub Copilot. In this scenario, the developer uses those tools to generate initial test scenarios, which are improved by the software developer. TDD iterations are carried out with the involvement of both the software developers and the LLMs, where developers supervise the results produced by the agent mode.

\textbf{Handing over Control to AI and Overreliance on AI \textcircled{8}} is evidenced when software developers overtrust AI tools for information retrieval, and do not also seek information from official documentation or other external reliable sources. By doing this, software developers are imprudently not taking into consideration the scenarios when AI generates content with hallucinations \cite{zhang:2025}. A similar issue arises when requesting LLMs to provide large-block code generation or code summarisation. Overall, these are scenarios where overrelying on AI tools, handing over the control to them, may impose negative consequences (e.g., degrade code quality). Nevertheless, software developers can effectively leverage LLMs to create prototypes, with the understanding that the software features will be revised based on user feedback. \textbf{Handing over Control to AI and Full Automation \textcircled{9}} is illustrated when software developers use AI tools to summarise main meeting points. Then, they can feel more comfortable interacting during the meeting. This also occurs when software developers let LLMs generate boilerplate code in their software projects.  

\textbf{Losing Control over AI and Overreliance on AI \textcircled{10}} is illustrated when software developers try to develop complete software solutions via LLMs, relying on their agent mode for also tasks like generating pull requests. They are start to lose their code understanding, and when some errors happen, they might need to continue relying on AI tools for debugging. \textbf{Losing Control over AI and Full Automation \textcircled{11}} is evidenced when software developers rely on LLMs' agent mode to conduct changes across multiple files. During this process, AIs can also modify critical sections of code, resulting in code vulnerabilities or errors. 

\section{Discussion} \label{sec:discussion}

\subsection{Implications for Practice}

\noindent \textbf{Appropriate tool selection: } Software team leaders can use our framework to help them with the selection of AI tools. For instance, they could select AI-driven development platforms requiring the level of control aligned with their team's expertise.  If the majority of the team consists of novice developers, software leaders should consider AI tools that offer an interaction mode requiring more control. At first glance, it may seem to reduce developers' productivity, but it can also ensure that code quality remains stable and hinders overreliance on AI.

 \textbf{Team management and policy:} Software team leaders can use the reliance-control framework to guide day-to-day management and establish norms for AI use. Teams can explicitly discuss and document when automation is appropriate and when human oversight is required, aligning levels of reliance and control with the criticality of the task. Regular reviews or retrospectives can include reflection on reliance patterns. Such discussions help surface unspoken practices, promote shared accountability for AI-generated artifacts, and reduce the risk of silent overreliance

\textbf{Guiding SE education AI tool usage: }SE educators can leverage the relation between control and reliance. Essentially, when introducing students to computing concepts, educators should allow students to use tools that require a high degree of human control, which may mitigate \textit{overreliance} and \textit{full automation}. For first- and second-year SE students, it is advisable to restrict their use of AI-assisted development environments when they are learning programming language syntax. These students should focus on understanding best software practices and should be limited to AI tools that demand active engagement from developers, such as conversational assistance. In contrast, final-year SE students should be introduced to software development using AI-driven platforms, as well as the potential risks of handing over control to AI. For instance, educators can employ role-play simulations to illustrate the limitations of LLM-powered tools, such as AI hallucination and context size limitation.

\subsection{Implications for Research} \label{sec:researchAgenda}


\noindent \textbf{Better understanding developer AI tool reliance in practice: } In our observations of various LLM-powered tools available in the market, we noted that different types of interactions offer varying levels of control for software developers. For example, using tools like ChatGPT and Claude allows developers to maintain a high degree of control (i.e., \textit{taking control}) over the context when communicating. In online platforms, developers must provide all contextual information, which can sometimes lead to ambiguity in expression. This issue is lessened when these tools are integrated directly into the development environment. Conversely, when developers use agent mode in AI-driven platforms such as Cursor and Windsurf, they relinquish some control to the LLMs. In agent mode, these tools can make changes across multiple files (i.e., \textit{handing over the control}), which increases the risk of LLMs inadvertently inserting errors or vulnerabilities into the code. This concern is highlighted by a large study \cite{qodo:2025} involving 609 software developers, which found that 25\% of participants believed that one in every five suggestions generated by LLMs may contain errors. Lastly, interaction with Claude Code in terminal mode represents a middle ground (i.e., \textit{balanced control}), allowing developers to engage with AI tools while still having the option to switch to standard terminal commands. This interaction suggests promoting an \textit{appropriate level of reliance} on AI. However, we recommend further investigation to validate these findings by conducting experiments with software developers using LLM-powered tools like Claude Code on a terminal.

\begin{table}
\scriptsize
\centering
\caption{Reliance-Control framework applied to the Developer-AI Interaction Types from  Treude and Gerosa \cite{treude:2025}.}
\label{tab:interactionModes}
\begin{tblr}{
  width = \linewidth,
  colspec = {Q[365]Q[333]Q[230]},
  row{odd} = {Gallery},
  hlines,
}
\textbf{Interaction Type}      & \textbf{Type of Control}       & \textbf{Type of Reliance}           \\
Conversational assistance      & Taking control over AI         & Appropriate reliance                \\
Selection-based enhancements   & Balanced control               & Appropriate reliance                \\
Explicit UI actions            & Taking control over AI         & Appropriate reliance                \\
Comment-guided prompts         & Balanced control               & Appropriate reliance                \\
Shortcut-activated prompts     & Taking control over AI         & Appropriate reliance                \\
Command-driven actions         & Taking control over AI         & Appropriate reliance                \\
Contextual recommendations     & Handing over the control to AI & {Overreliance on AI -\\Full Automation}  \\
Auto-complete code suggestions & Handing over the control to AI & {Overreliance on AI - \\Full Automation} \\
File-aware suggestion          & Handing over the control to AI & {Overreliance on AI -\\Full Automation}  \\
Event-based triggers           & Handing over the control to AI & Full Automation                          \\
Automated API responses        & Handing over the control to AI & Full Automation                          
\end{tblr}
\end{table}

 \textbf{Better understanding of the impact of developer expertise on AI tool selection and reliance: } To demonstrate an application of the framework to emerging knowledge in this area, we categorise the twelve developer-AI interaction types identified by Treude and Gerosa \cite{treude:2025} (see Table \ref{tab:interactionModes}). For example, auto-complete code suggestions and file-aware suggestion interaction types may be considered \textit{overreliance} while selection-based enhancements and comment-guided prompts interaction types can be consider \textit{appropriate reliance}.
 Although we did not consider development expertise in our emerging two-dimensional framework, literature presents expertise as one of the human factors influencing attitudes towards AI systems \cite{inkpen:2023, he:2023}. Experienced software developers generally feel more comfortable interacting with AI tools through the command line compared to novice developers. However, this method of interaction presents an opportunity to help novice developers become more familiar with working in a command-line environment. Further investigations could explore how the level of development expertise affects control and reliance on AI, improving our framework.


\textbf{Software Development Tasks and Appropriate Level of Reliance: } As we mentioned in Section \ref{sec:preliminaryfindings}, the level of control and reliance may differ according to the software development task, but also due to the capabilities of the AI tools. For instance, current AI-powered tools excel at repetitive tasks, such as code translation to other programming languages, which allows for a lower level of required oversight. We suggest future research to investigate the appropriate level of reliance on AI throughout the software development lifecycle.

 \textbf{Varying Interaction Modes according to Developer Expertise:} Developer expertise may also influence the level of reliance on AI-driven platforms. We suggested that novice developers use interaction modes that require users to \textit{take control}. In this context, these platforms can customise the user experience to align with developers' levels of expertise. However, it is crucial to strike a balance between control, reliance, and the productivity of software developers, since developers may not adopt tools that only slow them down. Therefore, further investigation is needed to determine how to achieve this balance effectively.

\section{Conclusion} \label{sec:impact}

Understanding the appropriate level of reliance on AI tools will allow software developers to take advantage of these tools for positive outcomes. This approach will not only benefit individual developers but also help organisations train software developers more quickly and with reduced risks associated with LLM adoption. By doing so, incidents like the one involving Replit-agent \cite{techco:2025} can be avoided or mitigated.

We propose a reliance-control framework through representative and practical examples. We recognise that the degree of control and reliance is not limited to the examples presented and discussed in this paper. We also recognise the potential research bias, common in qualitative research, and our limited sample size. As future investigation, we recommend validating and expanding these findings through case studies and experiments with software developers to better understand the role of control in LLMs for SE. Additionally, we suggest future work to compare different AI tools with varying supported interactions, in order to assess their impact on reliance in a more comprehensive manner.

\bibliographystyle{ACM-Reference-Format}
\bibliography{references}


\end{document}